\documentclass[12pt,twoside]{article}

\usepackage{amsmath}
\input BoxedEPS
\SetRokickiEPSFSpecial  %% for dvips by Tom Rokicki, for VMS
\HideDisplacementBoxes

\makeatletter
\long\def\@makecaption#1#2{%
  \vskip\abovecaptionskip
  \sbox\@tempboxa{\small #1: #2}%
  \ifdim \wd\@tempboxa >\hsize
   \small #1: #2\par
  \else
    \global \@minipagefalse
    \hb@xt@\hsize{\hfil\box\@tempboxa\hfil}%
  \fi
  \vskip\belowcaptionskip}
\makeatother

\newcommand{\bra}[1]{\bigl\langle#1\bigl|}
\newcommand{\ket}[1]{\bigr|#1\bigr\rangle}
\newcommand{\znn}{z_1,\ldots,z_n}
\newcommand{\zn}{z_1,\ldots,z_n}

\newcommand{\h}{Hamiltonian}
\newcommand{\hc}{h_{\mathrm C}}
\newcommand{\hp}{H_{\mathrm P}}
\newcommand{\hb}{H_{\mathrm B}}
\newcommand{\hn}[1]{H_{\mathrm{#1}}}
\newcommand{\tH}{\tilde H}
\newcommand{\Ht}{H(t)}
\newcommand{\he}{H_{\mathrm{E}}}
\newcommand{\aC}{A_{\mathrm{C}}}
\newcommand{\hee}[1]{H_{\mathrm{E,#1}}}
\newcommand{\ve}{V_{\mathrm{E}}}
\newcommand{\Hn}[1]{H(#1)}

\newcommand{\fract}[2]{{\textstyle\frac{#1}{#2}}}

\newcommand{\aea}{adiabatic evolution algorithm}
\newcommand{\ada}{adiabatic algorithm}
\newcommand{\q}{\quad}
\newcommand{\qq}{\qquad}
\newcommand{\qa}{quantum algorithm}
\newcommand{\qaa}{quantum adiabatic algorithm}

\newcommand{\qaea}{quantum adiabatic evolution algorithm}

\newcommand{\gs}{ground state}

\newcommand{\numeq}[2]{\begin{equation}
#2
\label{#1}
\end{equation}}
\newcommand{\refeq}[1]{(\ref{#1})}

\newcommand{\citer}[1]{~\cite{#1}}

\let\epsilon\varepsilon
\let\phi\varphi
\let\tilde\widetilde

\begin{document}
\title{Quantum Adiabatic Evolution Algorithms\\ 
 with Different Paths}
\author{Edward Farhi, Jeffrey
Goldstone\footnote{\tt farhi@mit.edu,
goldston@mit.edu}\\[-.75ex]
\footnotesize\itshape Center for Theoretical Physics, Massachusetts
Institute of Technology, Cambridge, MA 02139
\\ 
Sam Gutmann\footnote{\tt sgutm@neu.edu}\\[-.75ex]
\footnotesize\itshape  Department of Mathematics, Northeastern University,    
 Boston, MA 02115\\[-1.5ex]}
\date{\footnotesize\sf MIT-CTP~\#3297 \qquad  quant-ph/0208135}
\maketitle
\pagestyle{myheadings}
\markboth{E. Farhi, J.  Goldstone,   S. Gutmann}{Quantum Adiabatic Evolution
Algorithms  with Different Paths}
\vspace*{-2.25pc}

\begin {abstract}\noindent
In quantum adiabatic evolution algorithms, the quantum computer follows the ground state
of a slowly varying Hamiltonian. The ground state of the initial Hamiltonian is easy to 
construct; the ground state of the final Hamiltonian encodes the solution of the computational
problem. These algorithms have generally been studied in the case where the ``straight line''
path from initial to final Hamiltonian is taken. But there is no reason not to try  paths
involving terms that are not linear combinations of the initial and final Hamiltonians. We give
several proposals for randomly generating new paths. Using one of these proposals, we 
convert an algorithmic failure into a success.
\end{abstract}

\vspace*{-1pc}

\thispagestyle{empty}

\setcounter{equation}{0}
\subsection{Introduction}
\label{sec1}

Quantum \aea s\citer{r1} are designed to keep the quantum computer in the ground
state of a slowly varying Hamiltonian $\Ht$. The initial Hamiltonian, $H_{\mathrm B} = \Hn0$, is
chosen so that its \gs\ is easy to construct. The \gs\ of the final \h\ $\hp = \Hn T$,
where $T$ is the running time of the algorithm, encodes the solution to the
computational problem at hand. A simple way to construct  the interpolating \h\
is to define
\numeq{e1}{
\Ht = \tilde H(t/T) \qq 0\le t\le  T
}
where 
\numeq{e2}{
\tilde H(s) = (1-s) \hb + s H_{\mathrm P} \qq 0\le s\le 1\ . 
}

But there is no reason to restrict attention to this path from $H_{\mathrm B}$ to $\hp$.
The possibility of varying the path was discussed in\citer{r3},\citer{r4}, and\citer{rn4}.  The
\ada\ will work taking any path $\tH(s)$ with $\tH(0) = H_{\mathrm B}$ and
$\tH(1)=\hp$ as long as
$T$ is much larger than $1/\textit{gap}^2$ where the \emph{gap} is the minimum
energy difference, as $s$ varies, between the ground and first excited states of
$\tH(s)$. In this paper we consider  paths of the form
\numeq{e3}{
\tilde H(s) = (1-s)  H_{\mathrm B} + s H_{\mathrm P} + s(1-s) \he \ . 
}
We view $\he$ as an extra piece of the \h\ that is turned off at the beginning and end
of the evolution. For a given instance of  a problem, that is, for a fixed $\hp$, we
imagine running the algorithm repeatedly with different choices of $\he$, possibly
chosen randomly.

In this paper we propose some choices for $\he$. We then apply one of these choices
to a problem discussed in\citer{r2} and show that the addition of $\he$ can change
the performance of the algorithm from unsuccessful to successful. 

\subsection{Examples of different paths}
\label{sec2}
We are most optimistic about the performance of \qaa s when they are applied to the
classical problem of finding the minimum of a ``local'' cost function. By this we mean
a function
$h(\zn)$
where each $z_i=0,1$ and 
\numeq{e4}{
 h = \sum_{\mathrm C} \hc
} 
and each $\hc$ is a nonnegative integer-valued function that depends only on a
few~$z_i$. 

The \h\ that governs the \qa\ is constructed as follows. 
For  each term $\hc$ we define an associated quantum operator $H_{P,C}$. Suppose,
for example,  that $\hc$ depends on three bits $i_{\mathrm C}$, $j_{\mathrm C}$, and
$k_{\mathrm C}$. Then we have
\numeq{e5}{
H_{P,C} \ket{\zn} = 
\hc(z_{i_{\mathrm C}},  z_{j_{\mathrm C}},  z_{k_{\mathrm C}}) \ket{\zn}  \ .
}
Accordingly we define the problem \h 
\numeq{e6}{
\hp = \sum_{\mathrm C} \hn{P,C}\ .
} 
For the beginning \h\ we define
\numeq{e7}{
\hn{B,C} = \fract12 (1-\sigma_x^{(i_{\mathrm C})}) +
\fract12 (1-\sigma_x^{(j_{\mathrm C})}) +
\fract12 (1-\sigma_x^{(k_{\mathrm C})})
}
where (again) $i_{\mathrm C}$, $j_{\mathrm C}$, and $k_{\mathrm C}$ are the three bits
appearing in $\hc$. Now
\numeq{e8}{
H_{\mathrm B} = \sum_{\mathrm C} \hn{B,C} 
} 
and using \refeq{e6} and \refeq{e7} we see that $\tilde H(s)$ from \refeq{e2} can be written
\numeq{e9}{
 \tilde H(s)  = \sum_{\mathrm C} \bigl[(1-s) \hn{B,C} + s\hn{P,C}\bigr]\ .
}
This is the form of $\tilde H(s)$ that was originally proposed in equations~(2.26) and (2.28) 
in\citer{r1}.

We now turn to $\he$. To preserve the $\sum_{\mathrm C}$ on the right-hand side
of~\refeq{e9},  we pick a term $\hee{C}$ for  each term~$\hc$. This $\hee{C}$ will only
depend on the bits involved in~$\hc$. We then have
\numeq{e10}{
\he = \sum_{\mathrm C} \hee{C} 
}
and $\tilde H(s)$ from \refeq{e3} is now
\numeq{e11}{
\tilde H(s) =  \sum_{\mathrm C}   \bigl[  (1-s)\hn{B,C} + 
s\hn{P,C} +
s(1-s) \hee{C} \bigr] \ .
}
Note that for each clause the coefficients $(1-s)$, $s$, and $s(1-s)$ could all be replaced by any
three smooth functions with the same corresponding values at $s=0$ and $s=1$, but we make
the clause-independent choices 
$(1-s)$, $s$, and $s(1-s)$ for simplicity. 
\goodbreak

Finally we must choose $\hee{C}$. Here are three proposals. 
\begin{enumerate}
\item[P1.]  \label{item1} For each term $\hc$ involving $b_{\mathrm C}$ bits let
$\aC$ be a randomly chosen
Hermitian matrix of size $2^{b_{\mathrm C}} \times 2^{b_{\mathrm C}}$.  We imagine that the
distribution of $\aC$ produces entries with magnitude of order unity. We also propose
setting the diagonal entries in each
$\aC$ to~$0$. Now $\hee C$ is defined by letting the matrix $\aC$ act on the $b_{\mathrm C}$
bits involved in $\hc$. The $0$ diagonal avoids confusing the effects of varying the path in the
quantum algorithm with  modifications of the classical cost function itself. See\citer{rn4}.

\item[P2.]  \label{item2} For this proposal we specialize to the case where each term
$\hc$ has the same form independent of~$C$. For the case of three bits this means
that there exists an
$h_3$ such that
\numeq{e12}{
\hc (z_{i_{\mathrm C}}, z_{j_{\mathrm C}},  z_{k_{\mathrm C}}) = h_3 (z_{i_{\mathrm C}},
z_{j_{\mathrm C}},  z_{k_{\mathrm C}})   }
for each~$C$. (An example is the NP-complete problem Exact Cover where 
$h_3(z, z', z'') = 1$ if $z+z'+z''\neq1$ and $0$~if $z+z'+z'' = 1$.) For simplicity we restrict to the
case where each $\hc$ depends on three bits. We propose generating a single (order unity)
random $8\times8$ Hermitian matrix~$A$ and then using this for each $\hee C$. 
Each $\hee C$ has the $8\times8$ matrix~$A$ acting on bits $i_{\mathrm C}$,
$j_{\mathrm C}$, and $k_{\mathrm C}$. Again, we propose choosing the diagonal entries
of~$A$ to be~$0$. 

\item[P3.]   \label{item3} Here we make a proposal for 3-SAT. An $n$-bit instance of
3-SAT is a collection of clauses. Each clause specifies which three bits are in the
clause. Each clause is True for seven of the eight assignments of the three bits and
False on one assignment, so there are eight different types of clauses. 
Accordingly, $\hc=1$ if
$(z_{i_{\mathrm C}}, z_{j_{\mathrm C}},  z_{k_{\mathrm C}})$ is the False assignment
and $\hc=0$ on the other seven assignments. Thus $h(\zn)$ is the
number of clauses violated by the string $\zn$ and a satisfying assignment exists iff
the minimum value of~$h$ is~$0$. 

We can view the eight types of 3-SAT clauses as variants of a single basic clause, say
the one for which $(0,0,0)$ is the False assignment. The other seven clauses are
related to the basic clause by negation of some (or all) of the bits. For this proposal for
$\he$ we start with a single $8\times8$ matrix~$A$ and associate it with the basic
clause. For the other seven clauses we have seven other $8\times8$ matrices. Each is
obtained from~$A$ by negating the same bits that are negated in the corresponding
clause. (Bit negation of~$A$ is achieved by conjugating $A$ with the $\sigma_x$
matrix acting on the bit.) $A$ might be chosen randomly but again we take the
diagonal entries to be~$0$. 
This proposal has the invariance property that if an instance of 3-SAT is changed by
replacing some bits by their negations in all clauses, the performance of the quantum
algorithm is identical. 

\end{enumerate}

\subsection{A different path can turn failure into success}
\label{sec3}

Here we reexamine an example given in\citer{r2} where the \qaea, evolving
according to \refeq{e2}, without an $\he$, was shown to require a time exponential in
the number of bits. We will now show that for $\he$  chosen randomly according to 
Proposal~P\ref{item2} above, the \qaa\ succeeds in polynomial time.

The example has a cost function~$h$,  of the form \refeq{e4}, where each
$\hc$ depends on three bits and is of the same form for each~$C$, that is,
equation
\refeq{e12} holds. In this example
\numeq{e15}
{
h_3 (z ,z',z'') = \begin{cases}
0 & z + z' + z'' = 0\\
3 & z + z' + z'' = 1\\
1 & z + z' + z'' = 2\\
1 & z + z' + z'' = 3 
\end{cases}\ .
}
The cost function includes a term for each set of three bits giving
\numeq{e16}
{
h(\znn) = \sum_{i<j<k} h_3(z_i,z_j,z_k)\ .
} 
Note that the global minimum of $h$ occurs at $z_1=0, z_2=0,\dots,z_n=0$ where $h$
is~$0$. 

The performance of the \qaa\ on this  highly symmetrized problem can be analyzed
asymptotically in the number of bits~$n$. Using the symmetry and
equations~\refeq{e5}, \refeq{e6}, \refeq{e15}, and \refeq{e16} we can write
\begin{multline} \label{e17}
H_{\mathrm P} = \frac 32 \Bigl(\frac n2 - S_z\Bigr)\Bigl(\frac n2 +
S_z\Bigr)\Bigl(\frac n2+S_z -1\Bigr)\\
+ \frac12 \Bigl(\frac n2 - S_z\Bigr)\Bigl(\frac n2 - S_z -1\Bigr)\Bigl(\frac
n2+S_z\Bigr)\\
+ \frac16 \Bigl(\frac n2 - S_z\Bigr)\Bigl(\frac n2 - S_z -1\Bigr)\Bigl(\frac
n2-S_z-2\Bigr)
\end{multline}
where $S_z$ is the $z$-component of the total spin:
\numeq{e18}{
S_z  = \fract12 \sum_{i=1}^n \sigma_z^{(i)}\ .
}
For $H_{\mathrm B}$  we use \refeq{e7} and \refeq{e8} to obtain
\numeq{e19}
{
H_{\mathrm B} = \Bigl(\begin{matrix} n-1\\2\end{matrix}\Bigr)\Bigl(\frac n2 -
S_x\Bigr) }
where
\numeq{e20}{
S_x  = \fract12 \sum_{i=1}^n \sigma_x^{(i)} 
}
and for later use
\numeq{e21}{
S_y  = \fract12 \sum_{i=1}^n \sigma_y^{(i)} \ .
}

We now turn to $\he$ generated by Proposal~P\ref{item2}, from a single $8\times8$
matrix~$A$ with $0$'s on the diagonal. By \refeq{e10} and the fact that all sets of
three bits are included in the sum we find that $\he$ is a cubic function of $S_x$,
$S_y$, and $S_z$ whose coefficients are linear combinations of the entries of~$A$. 
The choice that $A$ is $0$ on the diagonal implies that there are no $S_z$, $S_z^2$,
and
$S_z^3$ terms in $\he$ whereas these are the only terms in~$H_{\mathrm P}$.

In this symmetric case the total \h\ $\tilde H(s)$ given by \refeq{e11} is a function of the total
spin operators $S_x$, $S_y$, and $S_z$. For $n$~large, we make 
the ansatz that for each~$s$ the ground state of $\tilde H(s)$ is a state
$\ket{\theta,\phi}$ that is an eigenstate of the total spin in the $(\theta,\phi)$
direction 
\numeq{e22}{
\bigl( \sin\theta\cos\phi\, \,  S_x +\sin\theta\sin\phi\, \,   S_y +\cos\theta\, \,   
S_z\bigr)
\ket{\theta,\phi} = \frac n2 \ket{\theta,\phi}\ .
}
With this ansatz finding the ground state of $\tilde H(s)$ is reduced to finding the
minimum over $(\theta,\phi)$ of 
\numeq{e23}{
\bra{\theta,\phi} \tilde H(s) \ket{\theta,\phi}\ .
}
To evaluate \refeq{e23} in the large-$n$ limit we use relations such as

\numeq{e24}{
\begin{aligned}
\bra{\theta,\phi} S_x \ket{\theta,\phi}  &= \frac n2 \sin\theta\cos\phi\\
\bra{\theta,\phi} S_y \ket{\theta,\phi}  &= \frac n2 \sin\theta\sin\phi\\
\bra{\theta,\phi} S_z \ket{\theta,\phi}  &= \frac n2 \cos\theta\\
\bra{\theta,\phi} S_x^2 \ket{\theta,\phi} &= \Bigl(\frac n2 \Bigr)^2 \sin^2\theta
\cos^2\phi + O(n)\\
\bra{\theta,\phi} S_x S_y \ket{\theta,\phi} &= \Bigl(\frac n2 \Bigr)^2 \sin^2\theta
\cos\phi\sin\phi + O(n)\\
\bra{\theta,\phi} S_x S_z \ket{\theta,\phi} &= \Bigl(\frac n2 \Bigr)^2 \sin\theta
\cos\theta\cos\phi + O(n)\\
    &\q \vdots
\end{aligned}
}
 We now define the effective potential 
\numeq{e25}{
V(\theta,\phi, s) = \lim_{n\to\infty} \Bigl(\frac2n\Bigr)^3
\bra{\theta,\phi} \tilde H(s) \ket{\theta,\phi}
}
which we break into two parts:
\numeq{e26}{
V_0 (\theta,\phi, s) = \lim_{n\to\infty} \Bigl(\frac2n\Bigr)^3
\bra{\theta,\phi} \bigl\{ (1-s)H_{\mathrm B} + s H_{\mathrm P}\bigr\}
\ket{\theta,\phi} } 
and
\numeq{e27}{
\ve (\theta,\phi, s) = \lim_{n\to\infty} \Bigl(\frac2n\Bigr)^3
\bra{\theta,\phi}   s(1-s) \he \ket{\theta,\phi}
}
so that 
\numeq{e28}{
V = V_0 + \ve\ .
}

\begin{figure}[ht]
\centerline{\BoxedEPSF{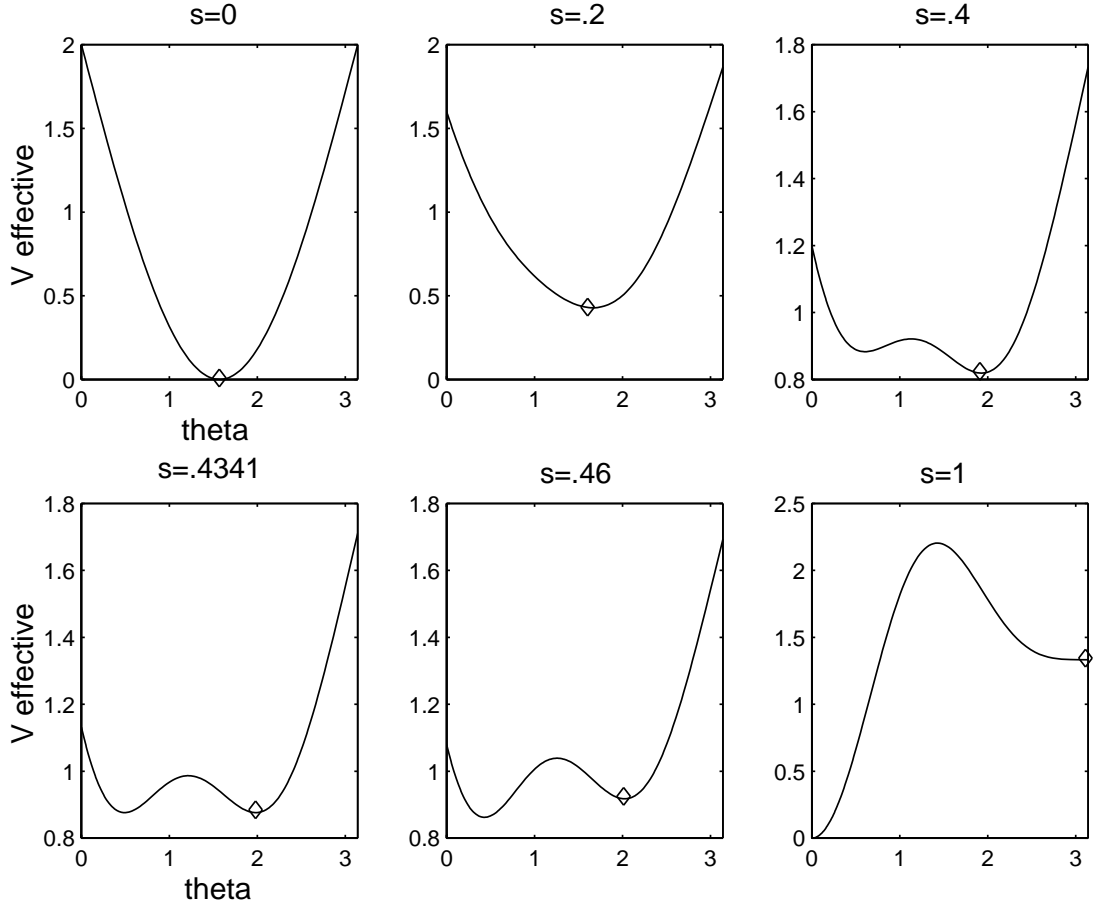 scaled 850}}
\caption{The diamond continuously tracks a local minimum of the effective potential,
corresponding to the behavior of the \qaa\ run for polynomial time. The final point reached
is not the global minimum of the effective potential at $s=1$ and the algorithm fails. Here
$\he$ is absent.}
\label{f1}
\end{figure}

We now review\citer{r2} what happens without $\he$. We have
\numeq{e29}{
V_0 (\theta,\phi, s) = 2 (1-s)  (1-\sin\theta\cos\phi) +
 \fract16 s (13 + 3\cos\theta-9\cos^2\theta -7\cos^3\theta)\ .
}
The minimum of $V_0$ for any $s$ is at $\phi=0$. In Figure~\ref{f1} we plot $V_0
(\theta,0, s)$ for six values of~$s$. In the first panel $s=0$ and the effective
potential is minimized at $\theta=\pi/2$, which corresponds to the ground state of
$H_{\mathrm B}$. As $s$ increases the minimum moves to the right (\emph{see} $s=.2$).
Continuously following this local minimum leads to the local minimum at $\theta=\pi$
when $s=1$. A quantum computer running the adiabatic algorithm for a time only
polynomial in~$n$ will evolve close to the state $\ket{\theta=\pi, \phi=0}$ at the end
of the run. This state $\ket{\theta=\pi, \phi=0} =\,\, \ket{z_1=1, z_2=1,\dots,z_n=1}$
does not correspond to the ground state of $H_{\mathrm P}$, which is at
$\ket{\theta=0, \phi=0} =\,\, \ket{z_1=0, z_2=0,\dots,z_n=0}$. 

The failure of the \qaa\ on this highly symmetrical problem with $n^3$ clauses may
not be a good test of the effectiveness of the algorithm on computationally interesting
problems. But in any event this failure can be turned to success with the addition
of~$\he$.

\begin{figure}[ht]
\centerline{\BoxedEPSF{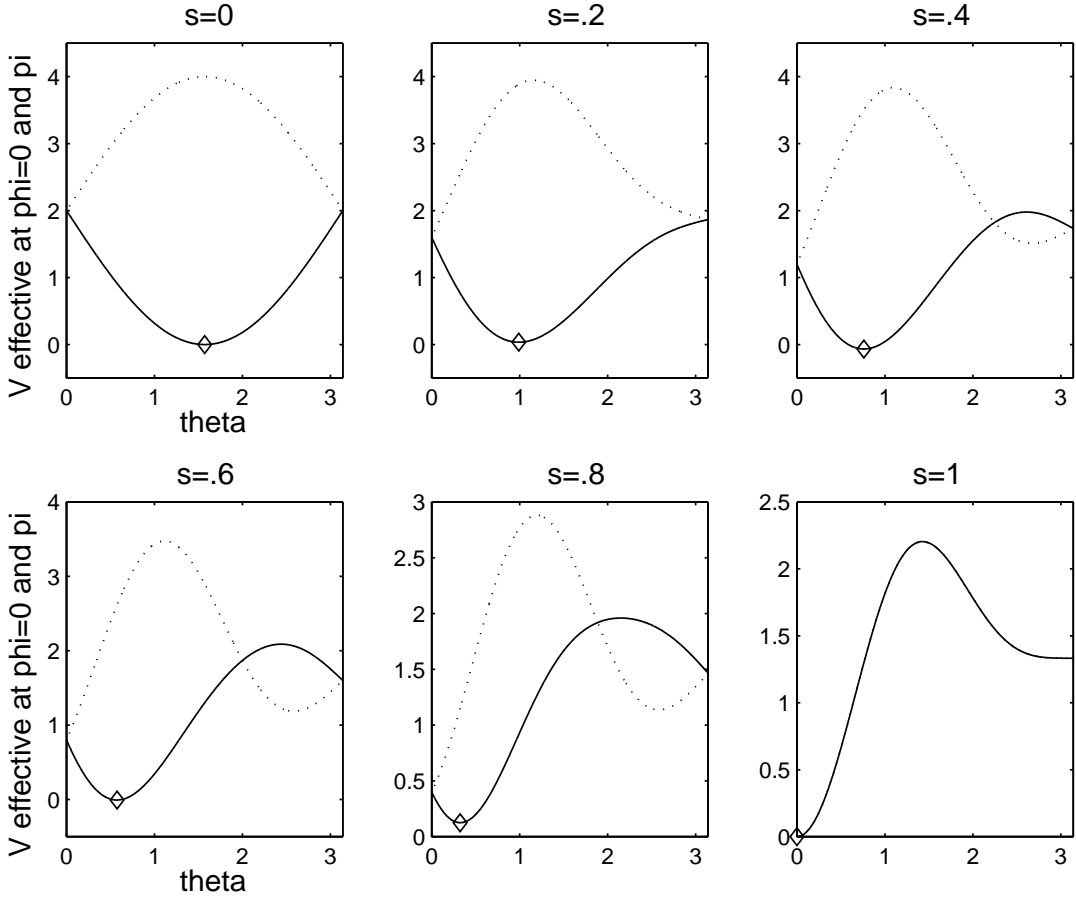 scaled 850}}
\caption{The diamond continuously tracks a local minimum of the effective potential for the
\h\ including a particular $\he$. (The solid line corresponds to $\phi=0$; the dotted line to
$\phi=\pi$.) The final point reached is the global minimum of the effective potential at $s=1$,
and the algorithm succeeds in polynomial time.}
\label{f2}
\end{figure}

Although Proposal~P\ref{item2} uses a random $8\times8$ matrix~$A$, we begin by
studying the effects of $\he$ arising from a carefully chosen~$A$:
\numeq{e30}{
A = \begin{bmatrix} 
0 & -2 &-2 &0 &-2 &0 &0 &0\\
-2 & 0 &0 &0 &0 &0 &0 &0\\
-2 & 0 &0 &0 &0 &0 &0 &0\\
0 & 0 &0 &0 &0 &0 &0 &2\\
-2 & 0 &0 &0 &0 &0 &0 &0\\
0 & 0 &0 &0 &0 &0 &0 &2\\
0 & 0 &0 &0 &0 &0 &0 &2\\
0 & 0 &0 &2 &0 &2 &2 &0
\end{bmatrix}
}
which gives rise to
\numeq{e31}{
\he = -2n(S_x S_z + S_zS_x) + O(n^2)
}
from which we get
\numeq{e32}{
\ve (\theta,\phi, s) = -8s(1-s)\cos\theta\sin\theta\cos\phi\ .
}
The full effective potential $V (\theta,\phi, s) $ given by \refeq{e28} with
\refeq{e29} and \refeq{e32} can only be extremized at $\phi=0$ and~$\pi$. 
 In Figure~\ref{f2} we show
  $V (\theta,0, s) $ as a solid line and  $V (\theta,\pi, s)$ as
a dashed line for six values of~$s$. At $s=0$ the ground state of $\tilde H(0)$
is $\ket{\theta=\pi/2, \phi=0}$.  In the first panel we see that $\theta=\pi/2,
\phi=0$ is the global minimum of $V (\theta,\phi, 0)$. As $s$ increases, the global
minimum moves continuously to $\theta=0, \phi=0$. The state
$\ket{\theta=0, \phi=0}$ $ =\,\, \ket{z_1=0, z_2=0,\dots,z_n=0}$ is the ground state of
$H_{\mathrm P}$ and corresponds to the minimum of the cost function given by \refeq{e15}
and \refeq{e16}. Running the \qaea\ for a  time polynomial in~$n$ will bring the
quantum computer close to the state $\ket{\theta=0, \phi=0}$,  solving the problem of
finding the minimum of~$h$.

The example just given is not an isolated success. Clearly the same behavior will be
seen for $A$'s close to \refeq{e30}. Moreover, we performed the following
experiment. We toss a random real symmetric $8\times8$ matrix~$A$ with
nondiagonal elements independently and uniformly distributed between~$-3$
and~$3$ and with the diagonal set to~$0$. We use~$A$ to construct $\he$ and the
corresponding $\ve$. 
We then continuously track a local minimum of $V (\theta,\phi, s)$ as $s$ varies from $0$
to~$1$, starting at $\theta=\pi/2,\phi=0$ when $s=0$. In 351 out of 1000 tries, this local
minimum moves to $\theta=0, \phi=0$, which is the global minimum of the effective potential
at~$s=1$. Thus, in these tries, the \ada\ run for polynomial time will end up close to the
ground state of $\hp$.  
 This shows that a large ``volume'' of the space of possible~$A$'s leads to an
$\he$ that ensures finding the global minimum of~$h$ in polynomial time. 

(It is conceivable that in some of these tries the continuously tracked local minimum, which
coincides with the global minimum 
at $s=0$ and $s=1$, does not coincide with the global minimum 
for all intermediate~$s$. If this happens, the algorithm
run for polynomial time will still succeed in finding the global minimum of~$h$, because the
probability of leaving the state corresponding to the local minimum is exponentially small.)

Our analysis, via the effective potential, of the \qa\ with an $\he$ chosen according to
Proposal~P2 gives the asymptotic (large~$n$) behavior of the algorithm. This analysis is only
possible because of the symmetry: the same terms are included in the \h\ for each set of
three bits. The function $h_3$ in \refeq{e15} could also be
viewed as arising from a sum of terms corresponding to thirteen 3-SAT clauses.
We could then use
Proposal~P3, which is designed for 3-SAT. Here one could apply the effective potential
analysis and determine the asymptotic behavior of the algorithm with random choices
of~$A$. Presumably the behavior is similar to  what we found using proposal~P2, but we
have not done the analysis.

 Proposal~P1 could be used instead in this problem, but choosing a different $\aC$
for each~$C$ breaks the symmetry and we don't know how to determine the large~$n$
performance of the algorithm in this case.  In\citer{rn4} a study of random 9-bit 3-SAT
instances was performed using proposal~P1. The gaps were compared with and without $\he$.
Overall the gaps with $\he$ were smaller than without~$\he$. However, the gaps that were
smallest without~$\he$ often increased after the addition of~$\he$ and were no longer small
compared to the other gaps. This argues for running the \qaa\ on each instance with a
variety of randomly chosen paths.
\vspace*{-\medskipamount}

\subsection{Conclusion}
\label{sec4}

Previous work on \qaea s has assumed a ``straight line'' path in \h\ space from the
beginning \h\ $H_{\mathrm B}$ to the problem \h\ $H_{\mathrm P}$ as in \refeq{e2}.
This path is simple but arbitrary. The idea of trying to remain close to the ground
state of a time-varying
\h\ is equally applicable to other paths from $H_{\mathrm B}$ to  $H_{\mathrm P}$,
such as \refeq{e3}. The
\qaa\ is guaranteed to give a 100\% probability of success only as the run time goes
to infinity, so  a finite run time requires repetition. 
We suggest that repetitions be carried out with different paths.
In cases where the gap in
$\tilde H(s)$ given by~\refeq{e2} happens to be small, the required run time is
correspondingly long. Adding $\he$ as in~\refeq{e3} may avoid the region in \h\
space where the gap is small and be computationally advantageous.

\subsubsection*{Acknowledgments}
We thank Michael Sipser and Wim van~Dam for useful discussions. 
This work was supported in part by the Department of Energy under cooperative
agreement DE-FC02-94ER40818 and by the National Security Agency (NSA) and
Advanced Research and Development Activity (ARDA) under Army Research
Office (ARO) contract DAAD19-01-1-0656. 

\vspace*{-\medskipamount}

\end{document}